\documentclass[%
tightenlines,%
 amssymb, amsmath,%
 aip,jap,%
]{revtex4-1}
\usepackage{graphicx}
\usepackage{dcolumn}%
\usepackage{array}
\usepackage{multirow}
\usepackage{overpic}
\usepackage{subfig}
\usepackage{bm}%
\usepackage[colorlinks=true,linkcolor=blue]{hyperref}%
\expandafter\ifx\csname package@font\endcsname\relax\else
 \expandafter\expandafter
 \expandafter\usepackage
 \expandafter\expandafter
 \expandafter{\csname package@font\endcsname}%
\fi
\newcommand{\super}[1]{\ensuremath{^{\textrm{#1}}}}
\newcommand{\sub}[1]{\ensuremath{_{\textrm{#1}}}}
\newcommand{\tilda}[1]{\ensuremath{\sim{}}}
\newcommand*{\citen}[1]{%
  \begingroup
    \romannumeral-`\x % remove space at the beginning of \setcitestyle
    \setcitestyle{numbers}%
    \cite{#1}%
  \endgroup   
}

\begin{document}

\title[]{On the diameter dependence of metal-nanowire Schottky barrier height}

\author{Yonatan Calahorra}\email{yoncal@tx.technion.ac.il}
\author{Eilam Yalon}
\author{Dan Ritter}\email{ritter@ee.technion.ac.il}

\affiliation{Dept. of Electr. Eng., Technion - Israel Inst. of Technology, Haifa 32000, Israel}

\begin{abstract}
Bardeen's model for the non-ideal metal-semiconductor interface was applied to metal-wrapped cylindrical nanowire systems; a significant effect of the nanowire diameter on the non-ideal Schottky barrier height was found. The calculations were performed by solving Poisson's equation in the nanowire, self-consistently with the constraints set by the non-ideal interface conditions; in these calculations the barrier height is obtained from the solution, and it is not a boundary condition for Poisson's equation. The main finding is that thin nanowires are expected to have tens of meV higher Schottky barriers compared to their thicker counterparts. What lies behind this effect is the electrostatic properties of metal-wrapped nanowires; in particular, since depletion charge is reduced with nanowire radius, the potential drop on the interfacial layer, is reduced - leading to the increase of the barrier height with nanowire radius reduction.
\end{abstract}
\maketitle
\section{Introduction}
Semiconducting nanowires (NWs) of a few to \tilda{}100 nm in diameter have been studied as future electronic-device building blocks, as well as active components in bio/chemical sensing systems, photovoltaics, photoemission, thermoelectrics and nano-electro-mechanical systems (NEMS) for the last 25 years \cite{Lu2007,Yang2014}. A prerequisite for successful and controllable device realization is a detailed understanding of the governing physics; in particular, understanding electrical contact properties is fundamental to device operation. When discussing nanomaterials, size and geometry may play a significant role in exhibited characteristics; e.g.\nolinebreak, axial (``end-on") contacts to nanowires were found to have diameter dependent properties \cite{Leonard2009,Leonard2011}.\\ \indent
Herein, the electrostatic analysis of metal-wrapped semiconducting NW contacts is presented, taking into account non-ideal interface effects, described by Bardeen's model \cite{Szebook}. The model introduces a thin interfacial layer between the metal and the semiconductor, and accounts for acceptor and donor like interfacial trap states; the barrier height is obtained from the solution of the problem instead of taken as a boundary condition for Poisson's equation. This distinction is crucial, since if the Schottky barrier is treated as a constant parameter \cite{Dobrokhotov2006,Park2010,Calahorra2013}, the electrostatic model does not allow incorporating the effect of surface states; indeed, theoretical studies dealing with the effect of surface states on the depletion properties of NWs, have considered the NW-ambient interface, rather than the NW-metal interface \cite{Schmidt2007,Simpkins2008,Chia2012}, which is of interest in many practical cases.\\ \indent
Since the resulting equations cannot be solved analytically, the barrier height as a function of NW radius is calculated numerically, and the results are discussed in an electrostatic context. The analysis shows that NW diameter has a significant effect on the barrier height, due to the size and geometry dependence of the problem's electrostatics.\\ \indent
\section{Theoretical background}
\subsection{Non-ideal metal-semiconductor interfaces}
\begin{figure}[hbt]
\centering
\includegraphics[scale=0.9]{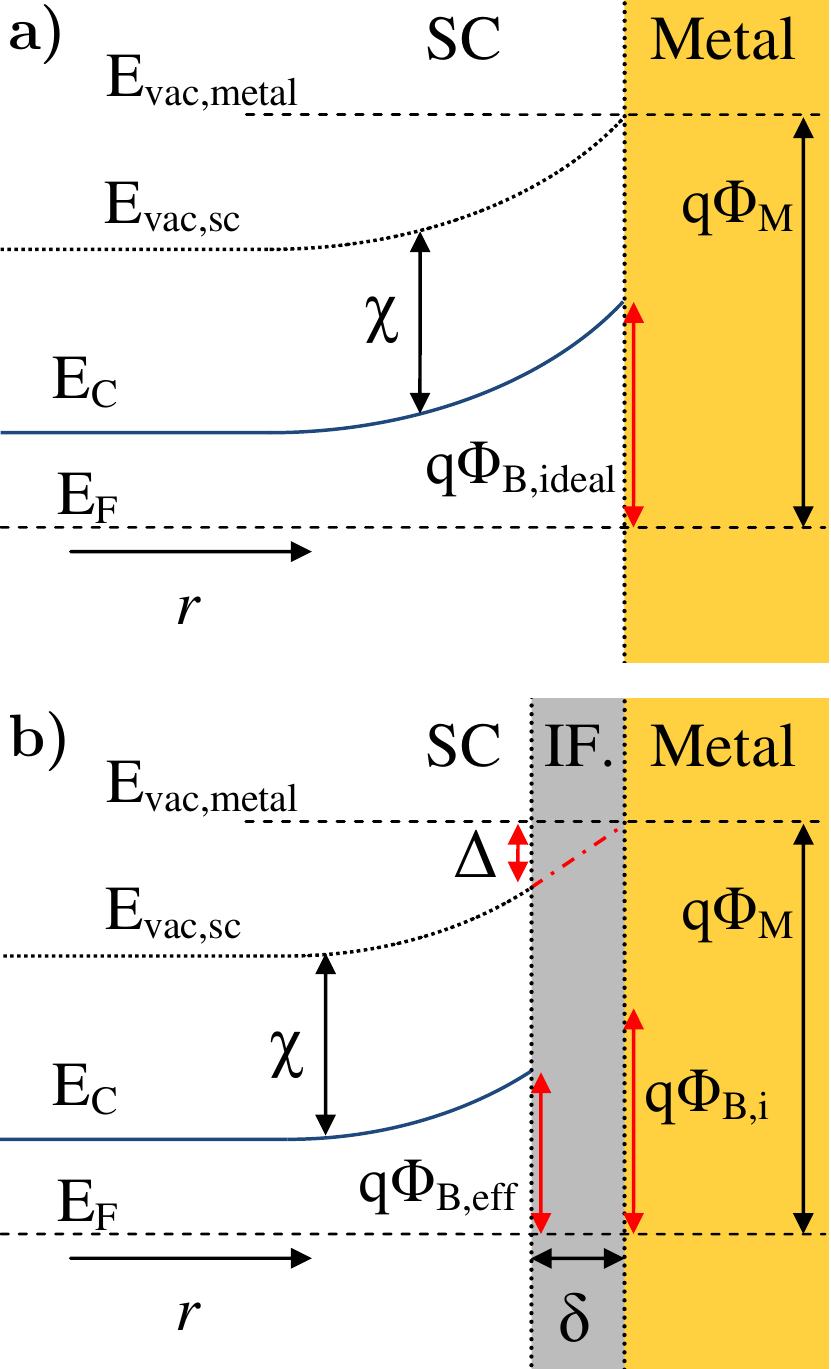}
\caption{Schematic representation of the metal-semiconductor contact; a) Ideal case; b) Non-ideal model, with the a voltage drop $\Delta$ on the interfacial layer.}  
\label{fig:scheme}
\end{figure}

The ideal model for the metal-semiconductor contact (Schottky-Mott rule) suggests that the Schottky barrier height is determined by the metal's work function, $\phi_m$, and the semiconductor electron affinity, $\chi$ by\cite{Szebook}
\begin{equation}
\phi_{B,ideal}=\phi_m-\chi
\label{eq:ideal}
\end{equation}
Figure \ref{fig:scheme} shows the schematic band diagrams of the metal-semiconductor system according to the ideal and non-ideal models. This ideal model was found to describe experimental results with only a limited degree of success, and a more elaborate model was introduced by Bardeen, accounting for the interface properties\cite{Szebook}; the essential relations for the planar case are reviewed here to allow the subsequent direct comparison to the cylindrical case.\\ \indent
In the non-ideal case (Fig.\nolinebreak\ \ref{fig:scheme}b), an interfacial layer of thickness $\delta$ lies between the metal and the semiconductor, and it may accommodate a voltage drop $\Delta$, which is the difference between the ideal barrier and the effective barrier
\begin{equation}
\Delta=\phi_{B,ideal}-\phi_{B,eff}
\label{eq:Delta1}
\end{equation}
furthermore, this voltage drop sustains the capacitive relations of a metal-oxide-semiconductor (MOS) system
\begin{equation}
\Delta=- \frac{\delta Q_{M}}{\epsilon_i} = - \frac{\delta \left( Q_{SS}+Q_{SC}\right) }{\epsilon_i}
\label{eq:Delta2}
\end{equation}
where $Q_{SS},Q_{SC},Q_{M}$ are the area charges of the surface, volume, and the metal, correspondingly. For simplicity, we assume the relative permittivity of the interfacial layer is 1, and $\epsilon_i$ is the vacuum permittivity.\\ \indent

In order to calculate the effective barrier the two expressions for $\Delta$ (Eq.\nolinebreak\ \ref{eq:Delta1},\ref{eq:Delta2}) are equated, and the charges are expressed with the effective barrier
\begin{eqnarray}\label{eq:Q_SS}
Q_{SS}&=&-qD_{it}\left( E_F-q\phi_0+E_V\right)=-qD_{it}\left( E_g-q\phi_0-q\phi_{B,eff}\right)\\\label{eq:Q_SC}
Q_{SC}&=&q W_{dep}(\phi_{B,eff}) N_D
\end{eqnarray}
where $\phi_0$ is the surface neutrality level. On the right-hand side of the equation, the difference between Fermi level and the neutrality level determines the surface charge, according to the product with the surface state density, $D_{it}$; states above (underneath) the neutrality level are acceptor (donor) type. The depletion charge is the depletion width times the doping concentration.\\ \indent
Combining Eq.\nolinebreak\ \ref{eq:Delta1}-\ref{eq:Q_SC} for the standard planar case and applying the depletion approximation, results in a simple quadratic equation; solving it yields the effective barrier height. Two physically distinct limiting cases exist, depending on the surface properties: 1) $D_{it}\longrightarrow 0$, where there is no surface charge and the barrier will approach the ideal value (if the depletion charge is neglected), or the barrier will be lowered according to the solution of the quadratic equation; 2) $D_{it}\longrightarrow \infty$, where surface charge dominates and Fermi-level pinning occurs. We set-out to examine how the application and results of this model would change, in cylindrical systems of various radii.\\ \indent
\subsection{Nanowire electrostatics}    
We consider a cylindrical semiconducting NW, wrapped by a metallic contact. A depleted shell will form at the NW surface, its thickness determined by NW doping; correspondingly, a radial band bending will occur, as shown schematically in Fig. \ref{fig:scheme}. This model is applicable to vertical NW devices, where the top contact encompasses the top section of the NW, and to omega-like side contacts, where most of the NW circumference is metal wrapped \cite{Lee2007,Lee2010,Gu2012,Hong2013,Bryllert2006,Bjork2010}.\\
Usually, in order to calculate the band diagrams, Poisson's equation in cylindrical coordinates is formulated with mixed boundary conditions, where, according to the depletion approximation, charge distribution results from ionized dopants
\begin{eqnarray}
\frac{d^2 \phi}{dr^2}+\frac{1}{r}\frac{d\phi}{dr}=-\frac{qN_D}{\epsilon_s}\hspace{20pt}  &;&\hspace{20pt}  R_{dep}<r<R \label{eq:NWPoisson}\\
\phi(R)=-\phi_{B}\  &;&\hspace{20pt}  \left.\frac{d\phi}{dr}\right\vert _{R_{dep}}=0 \label{eq:NWboundary}
\end{eqnarray}
where $\phi(r)$ is the potential, $N_D$ is the doping concentration (donors), and the conduction band, $E_C$, satisfies $E_C-E_F=-q\phi$. The solution and subsequent analysis have been presented previously \cite{Simpkins2008,ParkXu2010,Chia2012}. Recently we have expanded the analysis by considering the effects of radially non-uniform doping in NWs (i.e.\nolinebreak\ $N_D=N_D(r)$)\cite{Calahorra2013}; the existence of such doping profile has been demonstrated experimentally by several authors \cite{Pareige2012,Rosenwaks2010,Lauhon2009,Allen2009,Garnett2009,Amit2013,Connell2012}.  However, up until now electrostatic-oriented studies have considered the barrier height as a boundary condition, and not as the quantity to be calculated.\\ \indent
As will be discussed below in detail, the main result following the solution of Eq.\nolinebreak\ \ref{eq:NWPoisson}, is the dependence of the depletion width on NW radius: as the NW becomes smaller, the depletion width increases. Intuitively this may be understood looking at the second term (from the left) of the equation, which is the addition to the equation in cylindrical coordinates: in order to sustain the equation $d\phi/dr$ must drop as $r$ coordinate decreases and $1/r$ increases. Below we explore how the solution to the radial equation affects the non-ideal model.\\ \indent
The boundary condition at the NW surface $(r=R)$, describes the ideal barrier of the metal-semiconductor interface, and it could also be applied to a known case of Fermi-level pinning. The solution to Eq.\nolinebreak\ \ref{eq:NWPoisson} is given by
\begin{equation}
\phi(R_{dep}<r<R)=-\phi_B -\frac{qN_{D}}{\epsilon_s}\left[  \frac{R_{dep}^2}{2}\ln\left(\frac{R}{r}\right)-\frac{R^2}{4}+\frac{r^2}{4}\right] \label{eq:phi_}
\end{equation}
with $R_{dep}$ as a parameter. In order to find the depletion width, $W_{dep}=R-R_{dep}$, we consider the built in potential across the depletion region which is determined by $-\phi_B$ at the surface, and $(E_C-E_F)/q$ in the NW core; see Ref.\nolinebreak\ \citen{Calahorra2013} for a comprehensive discussion.\\ \indent
\section{Theoretical procedure}
\subsection{Applying Bardeen's model to nanowire contacts}
Stating Eq.\nolinebreak\ \ref{eq:Delta2} in cylindrical coordinates, requires expressing charge and capacitance per unit-length\cite{foot1}. The capacitance in this case is given by
\begin{equation}
C_{cyl}=\frac{2 \pi \epsilon_i}{ln \left( 1+\delta /R_{NW} \right) } \hspace{10 pt} \left[F/cm\right]
\label{eq:capac}
\end{equation}
and the equation is written
\begin{equation}
\Delta= - \frac{\left( Q_{SS,cyl}+Q_{SC,cyl}\right) }{C_{cyl}}
\label{eq:Delta2L}
\end{equation}
where
\begin{eqnarray}\label{eq:Q_SSL}
Q_{SS,cyl}&=&2\pi R_{NW}Q_{SS} \hspace{50 pt} \left[C/cm\right]\\\label{eq:Q_SCL} 
Q_{SC,cyl}&=&2\pi q \int\limits_0^{R_{NW}} r \cdot \left( N_D(r)-n(r) \right) \rm{dr} \hspace{10 pt} \left[C/cm\right]
\end{eqnarray}
it is important to recall that there is no closed form expression for the depletion width in cylindrical coordinates; moreover, when doping is non-uniform the problem's complexity increases. Therefore, in order to increase the solution's accuracy, the most general expression for the volume charge, $q\left( N_D(r)-n(r) \right)$, is used in Eq.\nolinebreak\ \ref{eq:Q_SCL}.\\ \indent
Under similar considerations Poisson's equation is written as
\begin{equation}
\frac{d^2 \phi}{dr^2}+\frac{1}{r}\frac{d\phi}{dr}=-\frac{q\left( N_D(r)-n(r) \right)}{\epsilon_s}\hspace{20pt}  \label{eq:NWPoissonL}
\end{equation}
where the potential at the interface gives the barrier, $\phi(R_{NW})=-\phi_{B,eff}$; it is crucial to acknowledge that the potential at the interface is \textit{not} a constraint or a boundary condition for the solution in this case, and that it is extracted upon solving the coupled equations. The potential and carrier concentration are related by Fermi-Dirac statistics\cite{Szebook}
\begin{equation}
n(r)=\frac{2}{\sqrt{\pi}}N_c F_{1/2}\left(q\phi/kT\right) \label{eq:FD}
\end{equation}
where $N_c$ is the effective density of states in the conduction band, and $F_{1/2}(x)$ is the Fermi-Dirac integral. We use an analytical approximation for the Fermi-Dirac integral, as presented by Bednarczyk and Bednarczyk\cite{Bednarczyk1978}, and applied by Chia and LaPierre in a NW related study \cite{Chia2012}. \\ \indent
A numerical procedure based on MATLAB's \textit{lsqnonlin} was developed and applied in order to solve the coupled equations \ref{eq:NWPoissonL} \& \ref{eq:FD} in the NW, under the constraints given by Eq.\nolinebreak\ \ref{eq:Delta1} \& \ref{eq:Delta2L} at the surface.\\ \indent
\subsection{Calculation details and model limitations}
The potential, surface and depletion charges, and the barrier, were calculated for three test cases: moderate, high, and a radial step-function doping 
\begin{eqnarray}
N_{D,1}&=&5\cdot 10^{16}\ \mathrm{cm^{-3}}\\
N_{D,2}&=&10^{18}\ \mathrm{cm^{-3}}\\
N_{D,3}(r)&=& \begin{cases} 6\cdot 10\super{18}\ \mathrm{cm^{-3}} & R_{NW}-r< 5\ \mathrm{nm} \\ 
10\super{18}\ \mathrm{cm^{-3}} &\mathrm{else} \end{cases} 
\end{eqnarray}
each of these cases was solved for four different surface conditions, one without surface states ($D_{it}=0$), and three with $D_{it}=10^{13}$ cm\super{-2}eV\super{-1}, and different locations of the neutrality level inside the band gap ($q\phi_0-E_V=0.4,\ 0.55,\ 0.7$ eV). The other parameters used are summarized in table \ref{tab:param}.\\ \indent

\begin{table}[hbp]
\caption{Fixed model parameters} 
\centering
\begin{tabular}{| c | c |}% c |}
\hline 
\textbf{Parameter} & \textbf{Value} \\ \hline \hline 
 R\sub{NW} [nm] & 15-200  \\ \hline 
Temp.\nolinebreak\ [K] & 300  \\ \hline 
$\epsilon_s$ [F/cm] & 12$\epsilon_0$ \\ \hline 
$\epsilon_i$ [F/cm] & $\epsilon_0$ \\ \hline 
$\delta$ [nm]& 0.5 \\ \hline 
$q\phi_{B,ideal}$ [eV]& 0.62 \\
\hline
\end{tabular} \label{tab:param}\\
%\textit{a} - analytical method induces $\pm5\%$ error\\
%\textit{b} - Result may be affected by AFM tip apex geometry\\
\end{table}
Two aspects of this analysis set the model's limitations: first, considering the diameter of the NWs, solving Poisson's equation rather than the coupled Schr\"odinger-Poisson equation, is valid only when the diameter is larger than the electron wavelength in the material. For standard semiconductors (such as silicon, GaAs, etc.\nolinebreak) this corresponds to 20-30 nm \cite{Mitin2008}; accordingly, the thinnest NWs considered hereunder are 30 nm in diameter. Second, the continuous model for the ionized dopant charge is valid when there are enough dopants such that a charge distribution function could effectively produce the results; when considering thin NWs, low dopant distributions could correspond to several or even a fraction of dopants per NW segment. In this work, the lowest concentration considered is $5\cdot10^{16}$ cm\super{-3}, which roughly corresponds to the few dopant limit per NW segment in 30 nm (diameter) NWs; the analysis, in large part, deals with a doping distribution of $10^{18}$ cm\super{-3}, so we consider the model applicable, and the results valid.\\ \indent 
\section{Results and discussion}
\begin{figure}[h]
\centering
\includegraphics[scale=1]{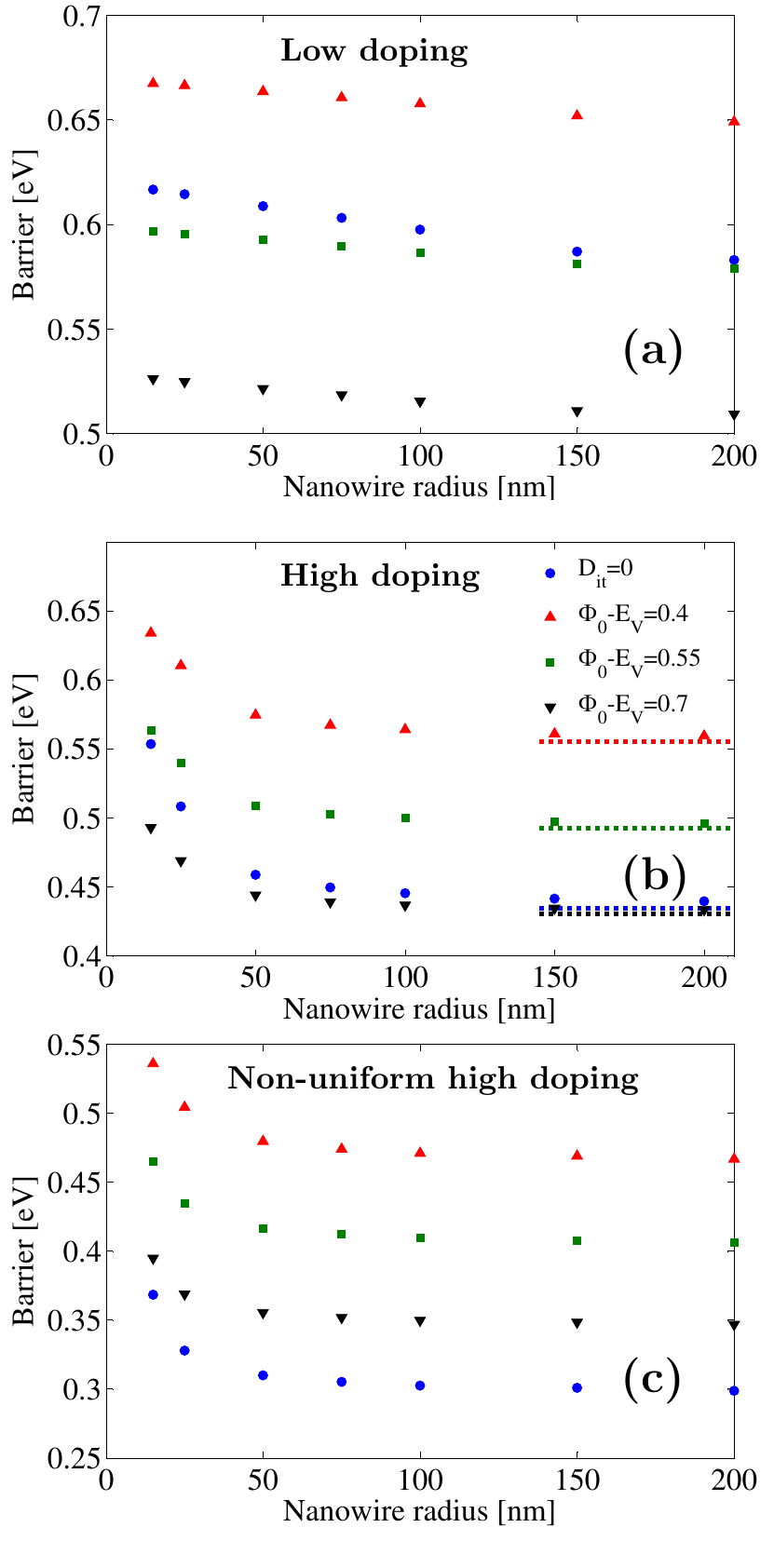}
\caption{Calculated non-ideal barrier height vs. NW radius: (a) low doping; (b) high doping; (c) non-uniform doping. In each sub-figure the barriers for different surface properties are shown, either no surface states (blue dots) or different locations of the neutrality level (triangles and squares as specified in the legend). The dashed lines in [b] represent the corresponding planar values.}  
\label{fig:bar}
\end{figure}

\begin{figure}[ht]
\centering
\includegraphics[scale=1]{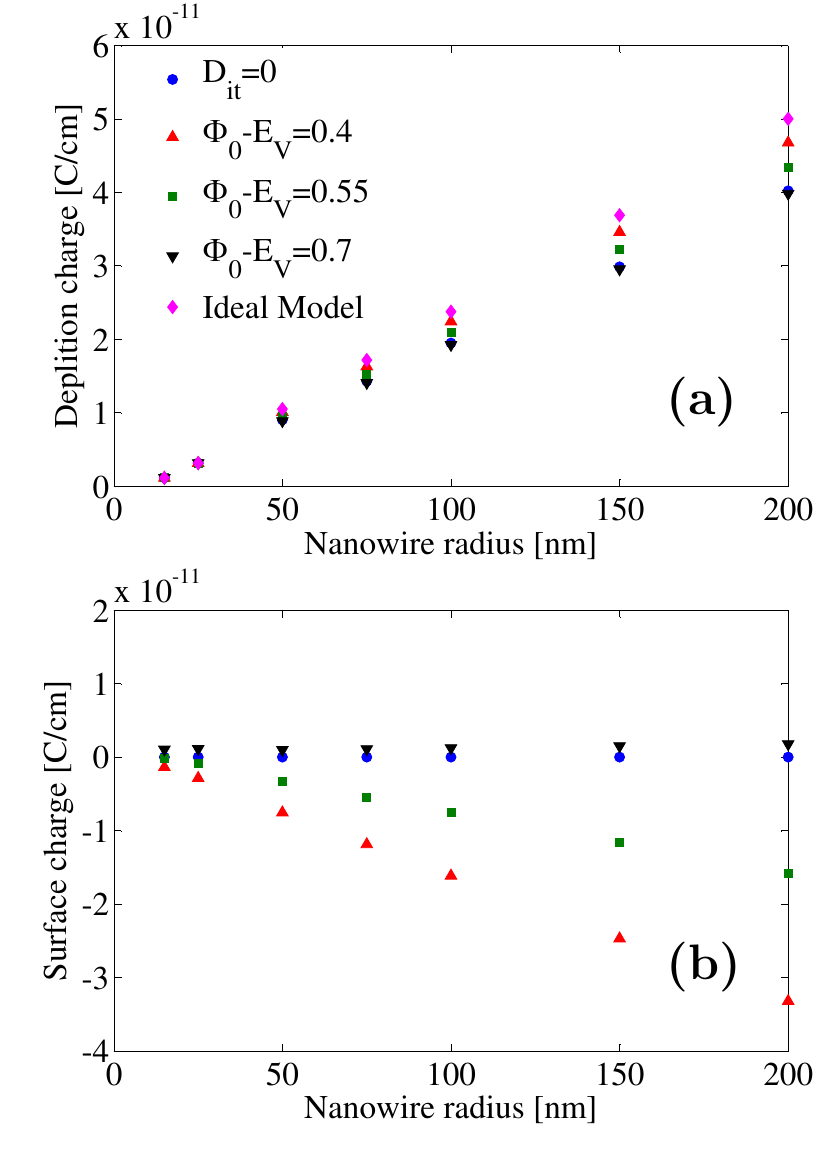}
\caption{Calculated (a) depletion and (b) surface charges per unit length for the high uniform doping case (Fig. \ref{fig:bar}b, data markers are the same). The depletion charge for the ideal model is added to [a] for comparison.}  
\label{fig:char}
\end{figure}

\begin{figure}[hbt]
\centering
\includegraphics[scale=1]{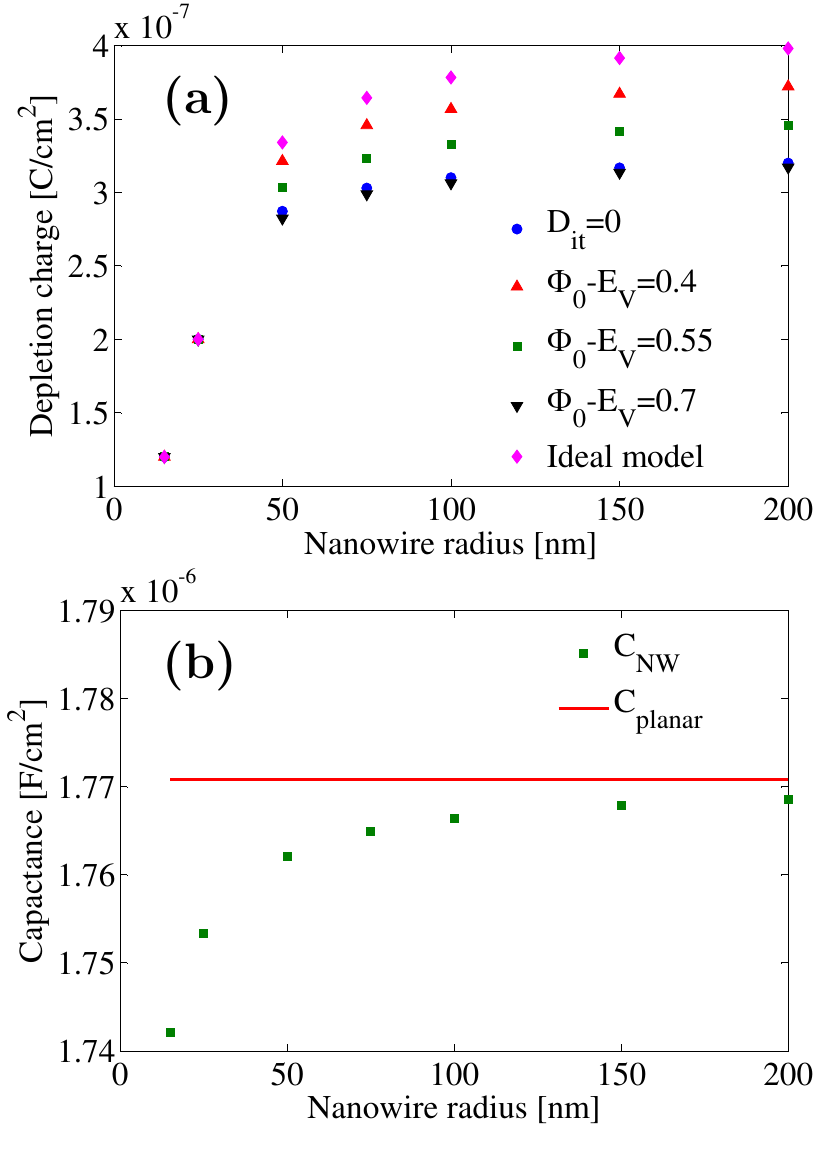}
\caption{(a) Calculated depletion charge per unit-area for the high doping case; this is the data shown in Fig. \ref{fig:char}a, normalized by the NW radius; (b) The capacitance per unit-area for the NW/interfacial-layer/metal system considered here as a function of NW radius, after Eq. \ref{eq:capac}; the planar value is shown for reference.}  
\label{fig:comp}
\end{figure}

Figure \ref{fig:bar} shows the calculated barrier height, as a function of NW radius, for the twelve cases under consideration (three doping profiles, four surface configurations). The most significant result, which is common to all simulation conditions, is the increase of barrier height with NW radius decrease; in some cases this is a very large effect (e.g.\nolinebreak \tilda{}100 meV for the surface state free case in highly doped NWs when going from a 50 to 15 nm NW), indicating that simply the size of NWs may have a significant effect on its contact properties. Two results were sought for, and act to validate the simulation results. First, the barrier height values for thick NWs, represented by the dashed lines in Fig.\nolinebreak\ \ref{fig:bar}b, which are in excellent agreement with the planar solution (Eq.\nolinebreak\ \ref{eq:Delta1}-\ref{eq:Q_SC}); second, the barrier height decrease with increased doping, exhibited in the difference between sub-figures.\\ \indent
Since the main result is exhibited in all cases, from now on we focus on the second case - high uniform doping; figure \ref{fig:char} shows the calculated surface and volume (depletion) charge for that case. The general trend of the NW depletion charge is clearly seen in Fig.\nolinebreak\ \ref{fig:char}a: even though the depletion shell width increases as NW radius decreases, the depletion charge decreases. For reference, we add the depletion charge calculated for the ideal case; it is clear that the trend of depletion charge reduction is a result of the problem's dimensionality and geometry, and not the application of the non-ideal model. The surface charge also decreases (in absolute value), however depending on the exact surface properties it can be either positive, negative or negligible. We note that the results are consistent even in the latter case, where there is no surface charge, indicating this is not a surface charge related effect. \\ \indent
In order to enhance the understanding of the governing physics, we offer a direct comparison to the more intuitive planar case; figure \ref{fig:comp}a shows the same results as in Fig.\nolinebreak\ \ref{fig:char}a, normalized by $2\pi R_{NW}$, yielding charge per unit-area values. The charge decrease trend is now further emphasized; moreover, as NW radius increases, the depletion charge (per unit-area) saturates, a result which is expected since large NW radii correspond to the planar case, where there should be no radial dependence.\\ \indent
In order to examine if the barrier dependence is merely a capacitive effect, we examine the influence of NW radius on the MOS capacitance. Figure \ref{fig:comp}b shows the cylindrical capacitance per unit-area, obtained according to Eq.\nolinebreak\ \ref{eq:capac} normalized by $2\pi R_{NW}$. The relative change in capacitance ($\tilde{}{1-2}\%$), is clearly not enough to account for the significant changes in the barrier hight, and in the depletion charge.\\ \indent
Considering the results for the barrier height and depletion charge, and the fact that the capacitance alone cannot account the observed effect, we suggest a simple qualitative explanation for the results, as follows. As NW radius is decreased, the depletion charge correspondingly decreases (Fig.\nolinebreak\ \ref{fig:char},\ref{fig:comp}); since the capacitance of the system hardly changes with radius, the voltage drop on the capacitor is reduced. On the other hand, the potential drop across the metal-semiconductor contact is geometry independent, and is given by the material parameters (Eq.\nolinebreak\ \ref{eq:ideal}); if so the rest of this potential drops across the depletion region, hence the barrier increases as the radius decreases.\\ \indent
The only experiments we are aware of, directly dealing with size effects on metal-NW barrier height are reported by Yoon et al.\nolinebreak\ \cite{Yoon2013}; contrary to our predictions, they have found a decreasing barrier height with NW size reduction, for N-type SiNWs. The theoretical model that Yoon et al. presented to account for their 
data, was not based on Bardeen's approach; we believe that Bardeen's approach is the appropriate model that should be applied, as in planar structures. However, a different effect may explain their results - related to doping incorporation. In chemical vapor deposition (CVD) NW growth, dopant incorporation through the NW sidewalls is a well-known and significant doping mechanism \cite{Lauhon2009,Pareige2012}; considering that NW sidewall area (where the dopant atoms enter the NW) down-scales with radius, and that NW volume (where the dopant atoms distribute) down-scales with the radius square - it could be that for given growth conditions, the doping concentration in thinner NWs is larger, what would result in lower Schottky barriers, as seen above according to the non-ideal model.\\ \indent
In a different study, by Kole{\'s}nik-Gray et al.\nolinebreak, researchers have observed the unpinning of Fermi-level at the surface of GeNWs \cite{Kolesnik2013}; this result points to the diminishing role of surface charge in the determination of contact properties, in agreement with our calculations. Further work is needed in order to determine to what extent their experimental results could be described by the model presented here, and what parameters could be extracted.\\ \indent
\section{Summary}
To conclude, we have rigorously addressed the issue of non-ideal metal-semiconductor interfaces in metal-wrapped NWs; this was done by applying Bardeen's model for the interface, to NW systems of various radii, doping profiles and surface properties. A numerical solution to the problem was calculated, and the results show a clear trend of increased barrier height with size reduction. This result intensifies with high doping levels, and interestingly, it is exhibited in the surface-state free case, as well as in the high surface state density cases. Importantly, the change of barrier height, calculated for values typical to silicon, can be as large as \tilda{}100 meV, which is expected to have a significant effect on NW device properties; in comparison, the image-force effect in NWs is smaller by 1-2 orders of magnitude \cite{Calahorra2014}. We show that the size dependence of NW depletion properties is behind this large effect: the depletion charge rapidly decreases with NW size reduction, and as a result the potential drop on the interfacial layer decreases as well. Subsequently, the remainder of the potential drops on the semiconductor depletion region. We believe that taking this effect into account in the analysis of NW devices with metal-wrapped Schottky contacts, will help in accurately determining important NW properties such as doping levels, and surface charge densities.\\ \indent
\section*{Acknowledgements}
The financial support of the Russell Berrie Nanotechnology Institute (RBNI), the Israel Ministry of Science and Technology (grant 38668) and the Israeli Nanotechnology Focal Technology Area on ``Nanophotonics for Detection", is thankfully acknowledged.
%

%\section*{References}
\bibliographystyle{unsrt}
\bibliography{size_barrier}{}

\begin{thebibliography}{10}

\bibitem{Lu2007}
Wei Lu and Charles~M Lieber.
\newblock Nanoelectronics from the bottom up.
\newblock {\em Nature materials}, 6(11):841--850, 2007.

\bibitem{Yang2014}
Neil~P Dasgupta, Jianwei Sun, Chong Liu, Sarah Brittman, Sean~C Andrews,
  Jongwoo Lim, Hanwei Gao, Ruoxue Yan, and Peidong Yang.
\newblock 25th anniversary article: semiconductor nanowires--synthesis,
  characterization, and applications.
\newblock {\em Advanced Materials}, 26(14):2137--2184, 2014.

\bibitem{Leonard2009}
Fran{\c{c}}ois L{\'e}onard, A~Alec Talin, BS~Swartzentruber, and ST~Picraux.
\newblock Diameter-dependent electronic transport properties of
  au-catalyst/ge-nanowire schottky diodes.
\newblock {\em Physical review letters}, 102(10):106805, 2009.

\bibitem{Leonard2011}
Fran{\c{c}}ois L{\'e}onard and A~Alec Talin.
\newblock Electrical contacts to one-and two-dimensional nanomaterials.
\newblock {\em Nature nanotechnology}, 6(12):773--783, 2011.

\bibitem{Szebook}
S.M. Sze and K.K. Ng.
\newblock {\em Physics of semiconductor devices}.
\newblock Wiley-interscience, 2006.

\bibitem{Dobrokhotov2006}
V~V Dobrokhotov, D~N McIlroy, M~G Norton, and C~A Berven.
\newblock Transport properties of hybrid nanoparticle–nanowire systems and
  their application to gas sensing.
\newblock {\em Nanotechnology}, 17(16):4135, 2006.

\bibitem{Park2010}
H.~Park, R.~Beresford, S.~Hong, and J.~Xu.
\newblock Geometry-and size-dependence of electrical properties of metal
  contacts on semiconducting nanowires.
\newblock {\em Journal of Applied Physics}, 108(9):094308--094308, 2010.

\bibitem{Calahorra2013}
Yonatan Calahorra and Dan Ritter.
\newblock Surface depletion effects in semiconducting nanowires having a
  non-uniform radial doping profile.
\newblock {\em Journal of Applied Physics}, 114(12):124310, 2013.

\bibitem{Schmidt2007}
V~Schmidt, S~Senz, and U~G{\"o}sele.
\newblock Influence of the si/sio2 interface on the charge carrier density of
  si nanowires.
\newblock {\em Applied Physics A}, 86(2):187--191, 2007.

\bibitem{Simpkins2008}
{Simpkins, BS and Mastro, MA and Eddy, CR and Pehrsson, PE}.
\newblock Surface depletion effects in semiconducting nanowires.
\newblock {\em Journal of Applied Physics}, 103(10):104313--104313, 2008.

\bibitem{Chia2012}
A.C.E. Chia and R.R. LaPierre.
\newblock {Analytical model of surface depletion in GaAs nanowires}.
\newblock {\em Journal of Applied Physics}, 112(6):063705--063705, 2012.

\bibitem{Lee2007}
S.Y. Lee and S.K. Lee.
\newblock {Current transport mechanism in a metal--GaN nanowire Schottky
  diode}.
\newblock {\em Nanotechnology}, 18(49):495701, 2007.

\bibitem{Lee2010}
S.N. Das, J.H. Choi, J.P. Kar, K.J. Moon, T.I. Lee, and J.M. Myoung.
\newblock {Junction properties of Au/ZnO single nanowire Schottky diode}.
\newblock {\em Applied Physics Letters}, 96(9):092111--092111, 2010.

\bibitem{Gu2012}
C.H. Hsu, Q.~Wang, X.~Tao, and Y.~Gu.
\newblock {Electrostatics and electrical transport in semiconductor nanowire
  Schottky diodes}.
\newblock {\em Applied Physics Letters}, 101(18):183103--183103, 2012.

\bibitem{Hong2013}
Young~Joon Hong, Chul-Ho Lee, Jun~Beom Park, Sung~Jin An, and Gyu-Chul Yi.
\newblock Gan nanowire/thin film vertical structure p--n junction
  light-emitting diodes.
\newblock {\em Applied Physics Letters}, 103(26):261116, 2013.

\bibitem{Bryllert2006}
T.~Bryllert, L.-E. Wernersson, L.E. Fr\"oberg, and L.~Samuelson.
\newblock {Vertical high-mobility wrap-gated InAs nanowire transistor}.
\newblock {\em Electron Device Letters, IEEE}, 27(5):323 -- 325, may 2006.

\bibitem{Bjork2010}
MT~Bj{\"o}rk, H~Schmid, CD~Bessire, KE~Moselund, H~Ghoneim, S~Karg,
  E~L{\"o}rtscher, and H~Riel.
\newblock Si--inas heterojunction esaki tunnel diodes with high current
  densities.
\newblock {\em Applied Physics Letters}, 97(16):163501, 2010.

\bibitem{ParkXu2010}
H.~Park, R.~Beresford, S.~Hong, and J.~Xu.
\newblock Geometry-and size-dependence of electrical properties of metal
  contacts on semiconducting nanowires.
\newblock {\em Journal of Applied Physics}, 108(9):094308--094308, 2010.

\bibitem{Pareige2012}
W.~Chen, V.G. Dubrovskii, X.~Liu, T.~Xu, R.~Lard{\'e}, J.~Philippe~Nys,
  B.~Grandidier, D.~Sti{\'e}venard, G.~Patriarche, and P.~Pareige.
\newblock {Boron distribution in the core of Si nanowire grown by chemical
  vapor deposition}.
\newblock {\em Journal of Applied Physics}, 111(9):094909, 2012.

\bibitem{Rosenwaks2010}
Elad Koren, Noel Berkovitch, and Yossi Rosenwaks.
\newblock {Measurement of Active Dopant Distribution and Diffusion in
  Individual Silicon Nanowires}.
\newblock {\em Nano Letters}, 10(4):1163--1167, 2010.
\newblock PMID: 20196550.

\bibitem{Lauhon2009}
D.E. Perea, E.R. Hemesath, E.J. Schwalbach, J.L. Lensch-Falk, P.W. Voorhees,
  and L.J. Lauhon.
\newblock {Direct measurement of dopant distribution in an individual
  vapour-liquid-solid nanowire}.
\newblock {\em Nature nanotechnology}, 4(5):315--319, 2009.

\bibitem{Allen2009}
Jonathan~E Allen, Daniel~E Perea, Eric~R Hemesath, and Lincoln~J Lauhon.
\newblock Nonuniform nanowire doping profiles revealed by quantitative scanning
  photocurrent microscopy.
\newblock {\em Advanced Materials}, 21(30):3067--3072, 2009.

\bibitem{Garnett2009}
Erik~C Garnett, Yu-Chih Tseng, Devesh~R Khanal, Junqiao Wu, Jeffrey Bokor, and
  Peidong Yang.
\newblock Dopant profiling and surface analysis of silicon nanowires using
  capacitance--voltage measurements.
\newblock {\em Nature Nanotechnology}, 4(5):311--314, 2009.

\bibitem{Amit2013}
Iddo Amit, Uri Givan, Justin~G Connell, Dennis~F Paul, John~S Hammond,
  Lincoln~J Lauhon, and Yossi Rosenwaks.
\newblock Spatially resolved correlation of active and total doping
  concentrations in vls grown nanowires.
\newblock {\em Nano letters}, 13(6):2598--2604, 2013.

\bibitem{Connell2012}
Justin~G Connell, KunHo Yoon, Daniel~E Perea, Edwin~J Schwalbach, Peter~W
  Voorhees, and Lincoln~J Lauhon.
\newblock Identification of an intrinsic source of doping inhomogeneity in
  vapor--liquid--solid-grown nanowires.
\newblock {\em Nano letters}, 13(1):199--206, 2012.

\bibitem{foot1}
The subscript \textit{cyl} denotes charge and capacitance per unit-length.

\bibitem{Bednarczyk1978}
D~Bednarczyk and J~Bednarczyk.
\newblock {The approximation of the Fermi-Dirac integral F\sub{1/2}($_\eta$)}.
\newblock {\em Physics letters A}, 64(4):409--410, 1978.

\bibitem{Mitin2008}
Vladimir~V Mitin, Viatcheslav~A Kochelap, and Michael~A Stroscio.
\newblock {\em Introduction to nanoelectronics: science, nanotechnology,
  engineering, and applications}.
\newblock Cambridge University Press, 2008.

\bibitem{Yoon2013}
KunHo Yoon, Jerome~K. Hyun, Justin~G. Connell, Iddo Amit, Yossi Rosenwaks, and
  Lincoln~J. Lauhon.
\newblock {Barrier Height Measurement of Metal Contacts to Si Nanowires Using
  Internal Photoemission of Hot Carriers}.
\newblock {\em Nano Letters}, 13(12):6183--6188, 2013.

\bibitem{Kolesnik2013}
Maria~M Kole{\'s}nik-Gray, Tarek Lutz, Gillian Collins, Subhajit Biswas,
  Justin~D Holmes, and Vojislav Krsti{\'c}.
\newblock Contact resistivity and suppression of fermi level pinning in
  side-contacted germanium nanowires.
\newblock {\em Applied Physics Letters}, 103(15):153101, 2013.

\bibitem{Calahorra2014}
Yonatan Calahorra, Dan Mendels, and Ariel Epstein.
\newblock Rigorous analysis of image force barrier lowering in bounded
  geometries: application to semiconducting nanowires.
\newblock {\em Nanotechnology}, 25(14):145203, 2014.

\end{thebibliography}

\end{document}